\documentclass[aip,rsi,amsmath,amssymb,preprint]{revtex4-1}

\usepackage{graphicx}
\usepackage{dcolumn}
\usepackage{color}
\usepackage{bm}

\newcommand{\un}[1]{\ensuremath{\,\mathrm{#1}}}

\begin{document}


\title{A 31\un{T} split-pair pulsed magnet for single crystal x-ray diffraction at low temperature}

\author{F. Duc}
\affiliation{Laboratoire National des Champs Magn\'etiques
Intenses, CNRS-INSA-UJF-UPS, 143, avenue de Rangueil, F--31400
Toulouse, France\\}
\author{X. Fabr\`{e}ges}
\affiliation{Laboratoire National des Champs Magn\'etiques
Intenses, CNRS-INSA-UJF-UPS, 143, avenue de Rangueil, F--31400
Toulouse, France\\}
\affiliation{Laboratoire L\'{e}on Brillouin, UMR12 CEA-CNRS
B\^{a}t 563 CEA Saclay, 91191 Gif sur Yvette Cedex, France\\}
\author{T. Roth}
\affiliation{European Synchrotron Radiation Facility, Bo\^{\i}te
Postale 220, F--38043 Grenoble Cedex, France\\}
\affiliation{European XFEL GmbH, Albert-Einstein-Ring 19, D-22761 Hamburg, Germany}
\author{C. Detlefs}
\affiliation{European Synchrotron Radiation Facility, Bo\^{\i}te
Postale 220, F--38043 Grenoble Cedex, France\\}
\author{P. Frings}
\affiliation{Laboratoire National des Champs Magn\'etiques
Intenses, CNRS-INSA-UJF-UPS, 143, avenue de Rangueil, F--31400
Toulouse, France\\}

\author{M. Nardone}
\affiliation{Laboratoire National des Champs Magn\'etiques
Intenses, CNRS-INSA-UJF-UPS, 143, avenue de Rangueil, F--31400
Toulouse, France\\}

\author{J. Billette}
\affiliation{Laboratoire National des Champs Magn\'etiques
Intenses, CNRS-INSA-UJF-UPS, 143, avenue de Rangueil, F--31400
Toulouse, France\\}

\author{M. Lesourd}
\affiliation{European Synchrotron Radiation Facility, Bo\^{\i}te
Postale 220, F--38043 Grenoble Cedex, France\\}

\author{L. Zhang}
\affiliation{European Synchrotron Radiation Facility, Bo\^{\i}te
Postale 220, F--38043 Grenoble Cedex, France\\}

\author{A. Zitouni}
\affiliation{Laboratoire National des Champs Magn\'etiques
Intenses, CNRS-INSA-UJF-UPS, 143, avenue de Rangueil, F--31400
Toulouse, France\\}

\author{P. Delescluse}
\affiliation{Laboratoire National des Champs Magn\'etiques
Intenses, CNRS-INSA-UJF-UPS, 143, avenue de Rangueil, F--31400
Toulouse, France\\}

\author{J. B\'{e}ard}
\affiliation{Laboratoire National des Champs Magn\'etiques
Intenses, CNRS-INSA-UJF-UPS, 143, avenue de Rangueil, F--31400
Toulouse, France\\}

\author{J.P. Nicolin}
\affiliation{Laboratoire National des Champs Magn\'etiques
Intenses, CNRS-INSA-UJF-UPS, 143, avenue de Rangueil, F--31400
Toulouse, France\\}

\author{G.~L.~J.~A. Rikken}
\affiliation{Laboratoire National des Champs Magn\'etiques
Intenses, CNRS-INSA-UJF-UPS, 143, avenue de Rangueil, F--31400
Toulouse, France\\}

\date{\today}

\begin{abstract}
We have developed a pulsed magnet system with panoramic access for
synchrotron x-ray diffraction in magnetic fields up to 31\un{T} and at
low temperature down to 1.5\un{K}. The apparatus consists of a split-pair
magnet, a liquid nitrogen bath to cool the pulsed coil, and a helium
cryostat allowing sample temperatures from 1.5 up to 250\un{K}. Using a
1.15\un{MJ} mobile generator, magnetic field pulses of 60\un{ms} length
were generated in the magnet, with a rise time of 16.5\un{ms} and a
repetition rate of 2 pulses/hour at 31\un{T}. The setup was validated for
single crystal diffraction on the ESRF beamline ID06.
\end{abstract}

\pacs{Valid PACS appear here}
\maketitle

\section{\label{sec:level1}Introduction}

Of all x-ray techniques, single crystal diffraction with a monochromatic
beam offers by far the highest resolution, the strongest absolute
signal, and the best signal to noise ratio. It is therefore the method
of choice for the examination of small crystallographic changes (such
as magneto-elastic effects), and for weak signals (such as magnetic
scattering or scattering of orbital order). Unlike x-ray diffraction
techniques such as white beam Laue diffraction or powder diffraction,
however, this technique requires precise alignment of the sample and
is therefore sensitive to vibrations that might be introduced by
electromagnetic forces induced by the field pulse. It requires optical
access over a wide angular range to enable measurements at high
scattering angles.

In many cases the magnetic field induced effects are anisotropic and
require the application of the field along a specific crystallographic
axis.  The challenge of combining x-ray single crystal diffraction and
pulsed high magnetic fields was taken up by several groups over the
world during the last ten years
\cite{Matsuda2004,Narumi2006a,Narumi2006,Matsuda2006,Islam2009,Islam2012}.

With this specific aim, the simplest way to obtain a coil with large
optical access in the mid plane perpendicular to the magnetic field
 is to use two coils, series
connected and axially aligned, with a gap between them. In this
arrangement, commonly called split-pair magnet, the vertical axial
bore is used to insert the sample cryostat, and the horizontal
mid plane gap is used for x-ray access. Over the years, several pulsed
magnets of split-pair type \cite{Matsuda2004, Narumi2006a, Narumi2006,
Matsuda2006, Islam2009} have been developed to meet the requirements
for combining x-ray diffraction with high pulsed magnetic field.
However, only very few publications \cite{Islam2012} have shown the
difficulties to associate both techniques and the efforts required to
reduce the pulse induced vibrations, and to make x-ray single crystal
diffraction compatible with pulsed magnetic field.

Here, a new pulsed field set-up, including a split-pair magnet and
sample cryostat, both specifically designed and optimized for single
crystal measurements on a synchrotron x-ray beamline, is
presented. Particular efforts have been carried out to quantify the
vibrations. Several technical changes have been made on the pulsed
magnet and cryogenic assembly by the LNCMI-Toulouse (LNCMI-T) and on
the beamline at the ESRF-Grenoble, to minimize the vibrations and
their propagation.

This device offers the possibility to invert the magnetic field
polarization and to change the sample temperature over a wide range
while optimally cooling the coil.  After a detailed description of the
different parts of the system we present its main features as regards
of field, temperature and angular resolution, illustrated by first results
obtained on a Fe$_{1.1}$Te single crystal.

\section{Pulsed field device and cryogenics}

\subsection{\label{sec:level2}The split-pair magnet: design and construction}

The split-coil magnet described here was designed and built at the
LNCMI-T. It is composed of two coils consisting of 36 layers of 13
turns.  The wire choice was the result of a trade-off between low
resistivity to reduce coil heating and high
ultimate tensile stress (UTS) to resist to magnetic pressure. A copper
conductor, internally reinforced by niobium titanium filaments and
insulated with kapton, was used. Its main features are reported in
Table~\ref{tab.wire}.

\begin{table}[h]
\caption[Wire]{\label{tab.wire}\small\sl{Main mechanical and
    electrical properties of the wire conductor. The copper wire was
    reinforced by 630 NbTi filaments of around 80~$\mu$m diameter. The
    volume fraction of the NbTi determined from the number of
    filaments and their average diameter is around $56.6\%$.}}
\begin{tabular}{p{5cm} l}
\hline
\hline\\
Cross section & 5.77 mm$^2$\\
Thickness $\times$ width & 2 mm $\times$ 3.15 mm \\
Ultimate tensile stress &
UTS$_{77K}\sim$ 1 GPa \\
Resistivity & $\rho_{77K}\sim$ 0.35 $\mu\Omega$.cm \\
& $\rho_{300K}\sim$ 2.9 $\mu\Omega$.cm \\
\hline
\hline
\end{tabular}
\end{table}

\begin{figure}[h]
{\centerline{\includegraphics[width=0.7\columnwidth,clip,angle=0]{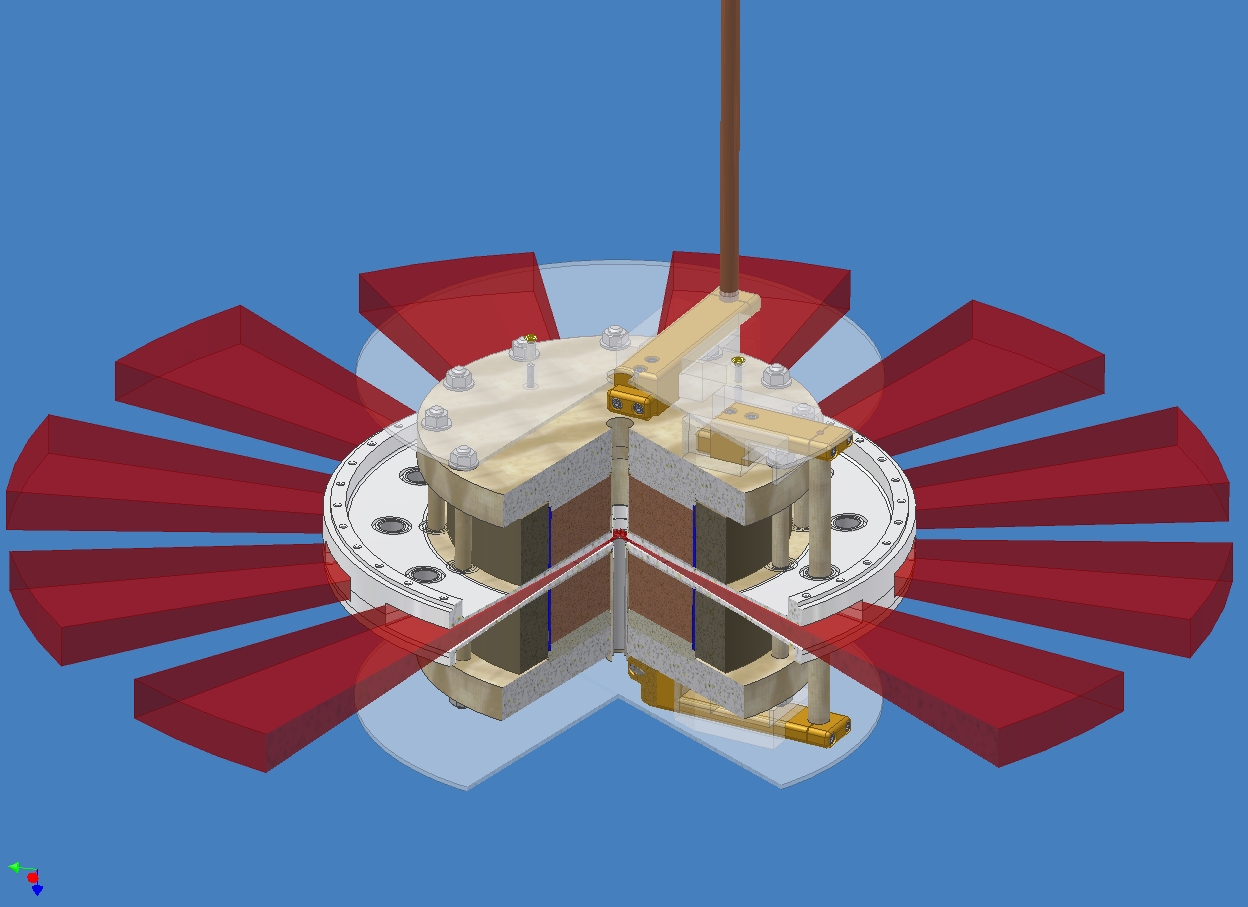}}}
\caption[Split and Beam] {\small\sl{
    (Color online) Three-dimensional view of the split-pair magnet. The
    optical access for x-ray beam in the mid-plane are highlighted by
    the red color.
}}\label{Split3Dview}
\end{figure}

The usable space for sample and cryogenics is determined by the magnet
bore diameter and the gap between the two coils.  Stainless steel
spacers were inserted between both magnets to withstand the pressure
($\sim$ 200~MPa) resulting from attraction induced by Lorentz forces
on the two coils.  These spacers were designed with $16^\circ$ open
sectors for x-ray beam access alternating with $14^\circ$ closed
sectors (see Fig.~\ref{Split3Dview}).  A thin layer of Teflon and
2\un{mm} of glass fiber epoxy composite (FR4/G10) were added between the
coils and the spacers to increase electrical insulation and internal
mechanical reinforcement.
\begin{figure}[h]
{\centerline{\includegraphics[width=1\columnwidth,clip,angle=0]{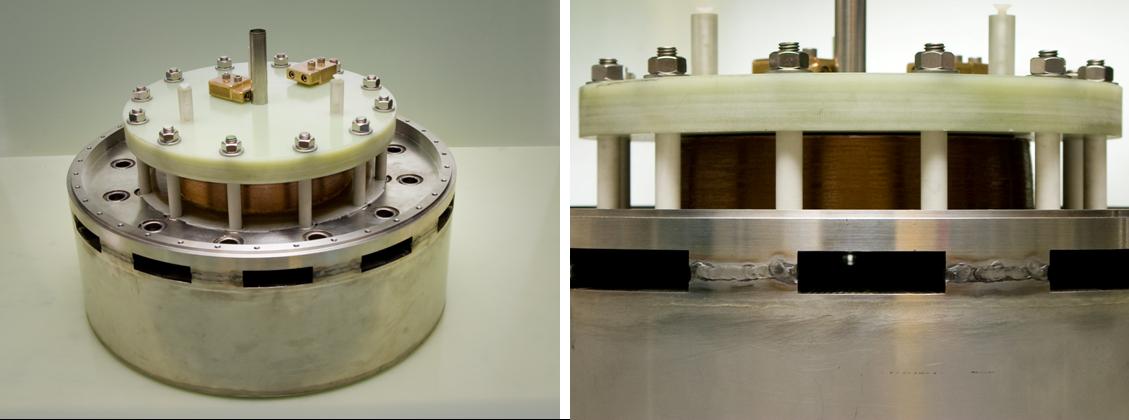}}}
\caption[Pictures of coil] {\small\sl{
    (Color online) Pictures of the split-pair magnet mounting.  Only the
    upper coil is visible, the lower coil being integrated into the
    lower part of the liquid nitrogen bath.
}}\label{Splitcoil}
\end{figure}

With an inner diameter of 22\un{mm}, an outer diameter of 200\un{mm},
a height of 50\un{mm} per coil and a gap of 20\un{mm}, the total
height of the split-pair magnet is 120\un{mm} and the resulting
inductance $L = 35\un{mH}$.  At the center of the magnet, the
vertical length available for the sample is of $\pm 2.5\un{mm}$ and the
vertical angular opening of each open sectors is of $\pm 2.5^\circ$
(see Fig.~\ref{Splitcoil}).

This split-pair magnet was designed to generate a maximum field of
30\un{T} at the sample position. This design goal was achieved as the
final coil reached a field of 31\un{T} in user operation.

\subsection{\label{sec:level2}Cryogenic environment description}
The cryogenic system consists of a liquid nitrogen (LN2) bath cryostat
that houses the coils, and a separate liquid $^4$He cryostat for the sample.
\subsubsection{\label{sec:level3}LN2 bath cryostat}
\begin{figure}[h]
{\centerline{\includegraphics[width=0.7\columnwidth,clip,angle=0]{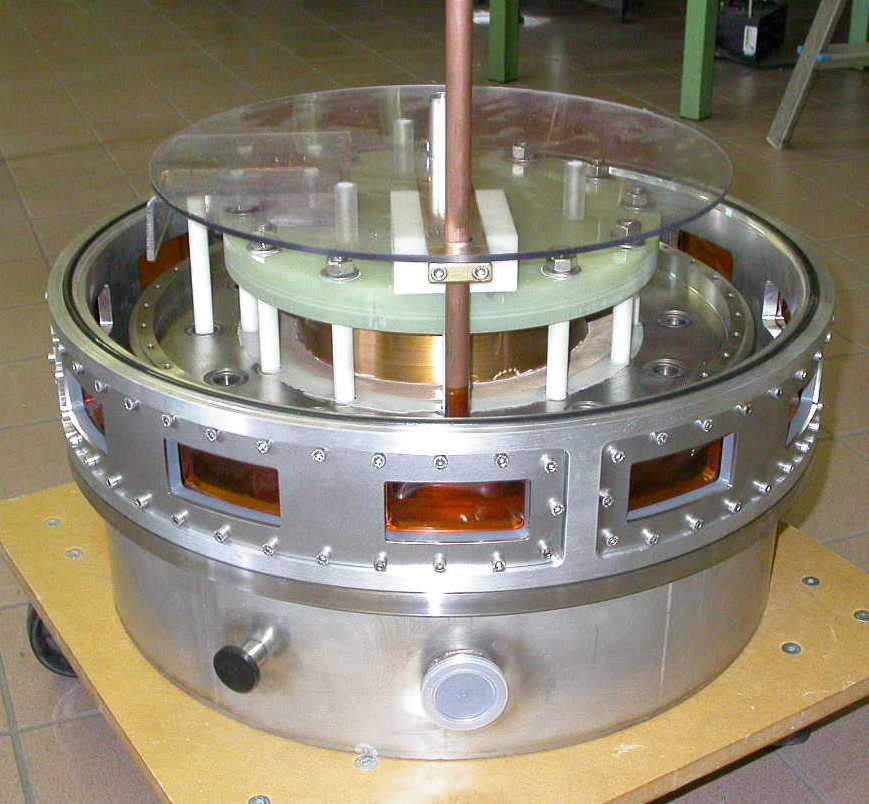}}}
\caption[Split mounting] {\small\sl{
    (Color online) Split-pair magnet in the lower part of the nitrogen
    cryostat. Kapton windows distributed around the dewar are visible.
}}\label{Splitmounting}
\end{figure}

The split-pair magnet was integrated into a custom designed nitrogen
cryostat equipped with twelve x-ray transparent, vacuum tight Kapton
windows (foils of 120\un{\mu m} thickness). The latter are distributed around the Dewar
(Fig.~\ref{Splitmounting}), matching the $16^\circ$ optical access
sectors in the spacers between the magnets. The internal vessel of this cryostat is
divided in two parts (the lower and upper baths), each containing a
coil (Fig.~\ref{Cryo3D}). A stepped tube, intended to house the
$^4$He insert, passes through the bore of the split-coil (see
Fig.~\ref{Splitcoil} and \ref{Cryo3D}). Connection and continuity
between the lower and upper baths are ensured by twelve stainless
steel tubes traversing the spacers. This configuration allows the
use of a common vacuum for the nitrogen cryostat, the helium insert
and the x-ray beam path.
\subsubsection{\label{sec:level3}$^4$He insert cryostat}
The coil/LN2 cryostat assembly hosts a top loading $^4$He cryostat allowing
measurements in the temperature range 1.5 -- 250~{K}.

\begin{figure}[h]
{\centerline{\includegraphics[width=\columnwidth,clip,angle=0]{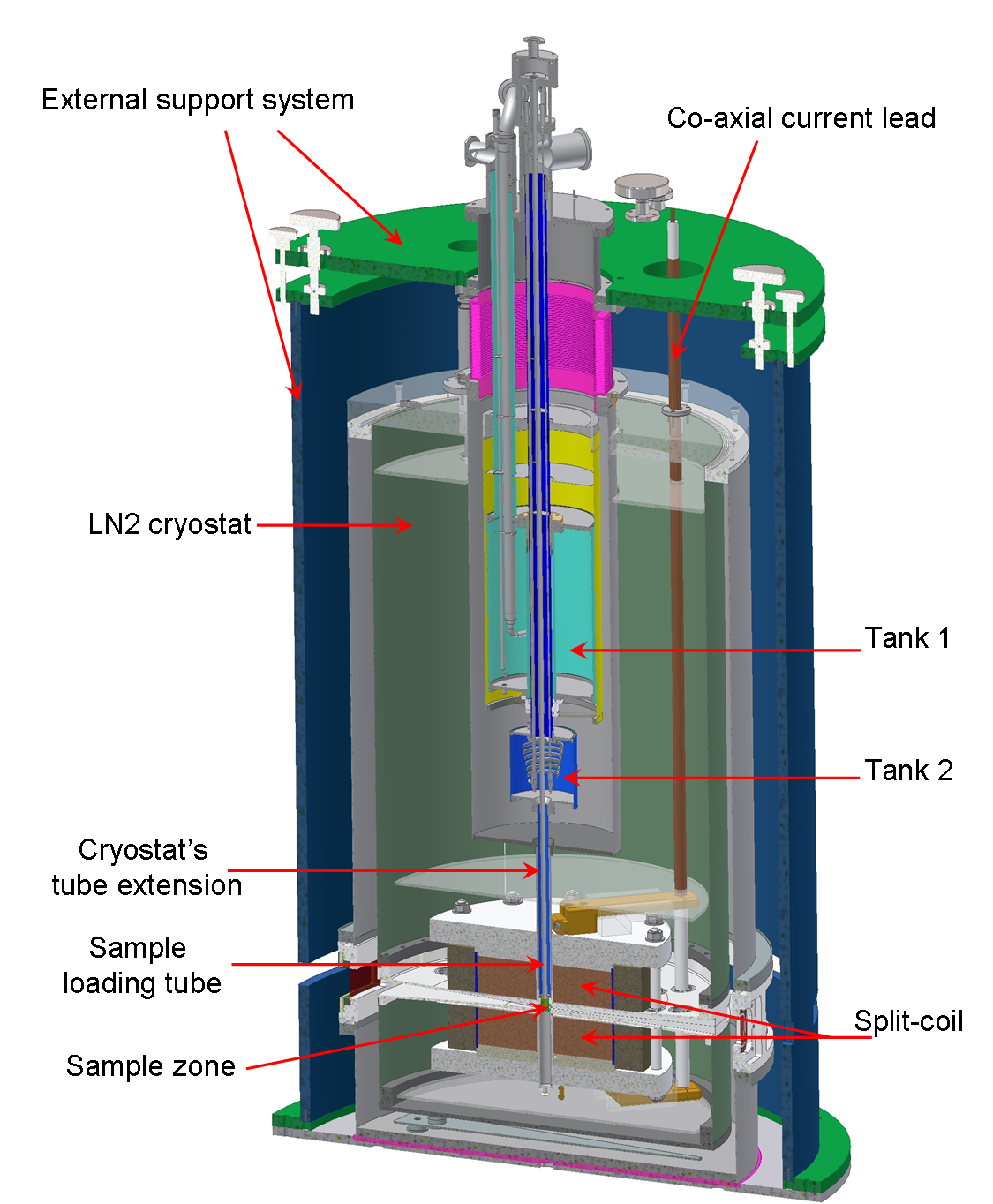}}}
\caption[3D cryogenics] {\small\sl{
    (Color online) Three-dimensional schematic view of the split-pair magnet and
    cryogenic environment.
}}\label{Cryo3D}
\end{figure}

\begin{figure}[h]
{\centerline{\includegraphics[width=0.9\columnwidth,clip,angle=0]{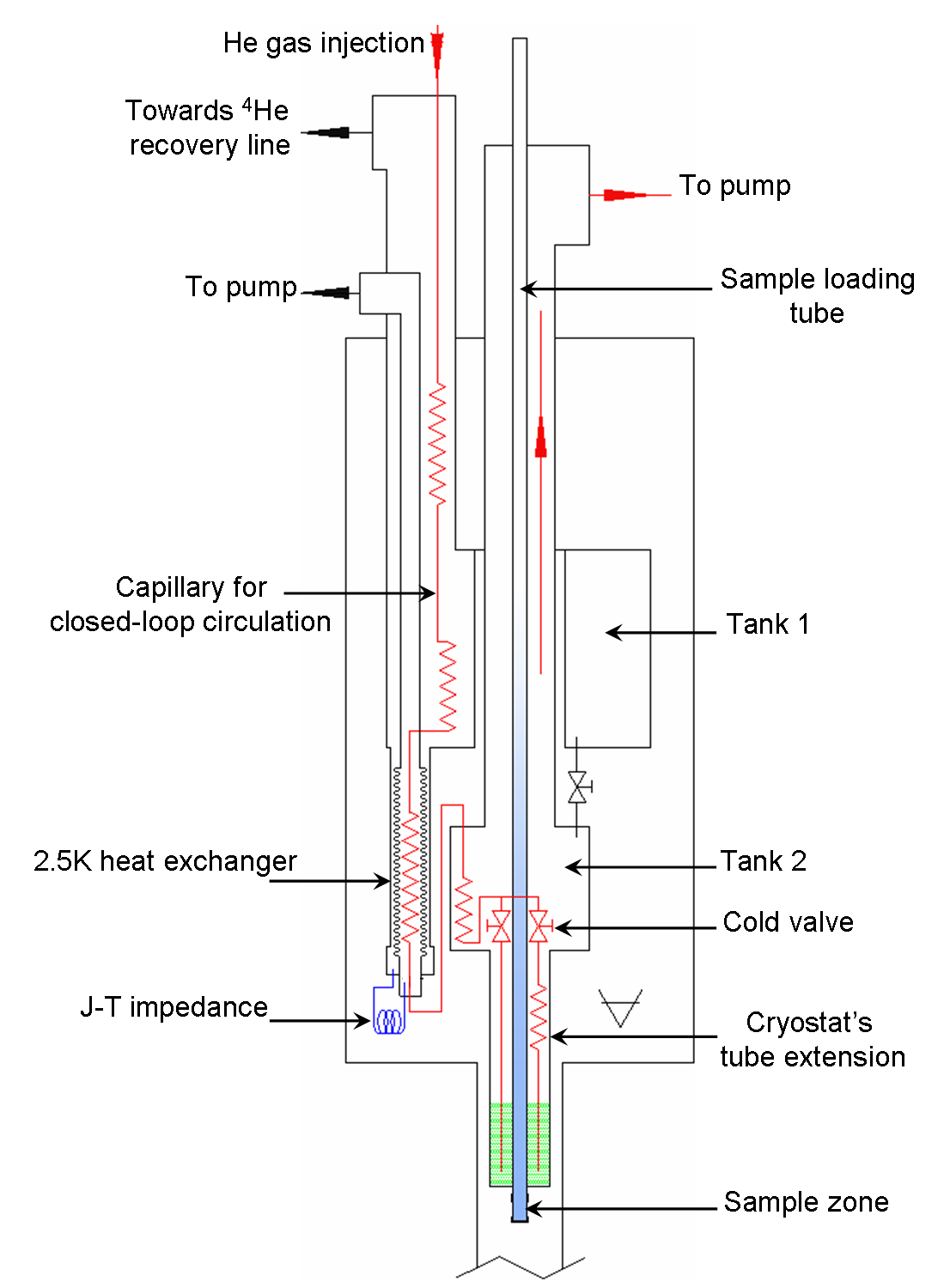}}}
\caption[3D cryogenics] {\small\sl{
    (Color online) Scheme of the $^4$He insert cryostat.
}}\label{Hecryostat}
\end{figure}

As shown on figures ~\ref{Cryo3D} and \ref{Hecryostat}, the $^4$He cryostat consists of five
parts: the main bath (tank 1), a 2.5\un{K} heat exchanger, the 1.5\un{K} bath
(tank 2) with its extension, the sample loading tube system and the
sample zone.

The main bath (tank 1) is a 3l liquid helium vessel. It
is connected with the 1.5\un{K} bath (tank 2, 0.4l) via a tube of 2\un{mm}
inner diameter equipped with a cold valve. When open, the valve
allows refilling of tank 2 with liquid $^4$He from the main bath
while, when closed, tank 1 can be refilled with almost no effect on the
temperature of tank 2. The latter is extended by a vertical stainless
steel tube of 15\un{mm} outer diameter (the ``cryostat's tube extension'')
equipped with a joule heater and a Cernox calibrated thermometer.

The cryostat includes also a 2.5\un{K} heat exchanger
pumped by an independent pumping line.
This exchanger is continuously supplied with liquid helium coming from the tank 1
and includes in its lower part a Joule Thomson (J-T) impedance (capillary
of 10\un{cm} length and 0.15\un{mm} inner diameter) allowing the expansion of the helium gas
and, therefore, the cooling of the vapors in the inner part of the exchanger down to 2.5\un{K}.
In closed cycle mode, the He flow is injected through a long capillary (stainless steel tube of 1.2\un{m}
length and 1\un{mm} inner diameter), wound inside the exhaust line of
the tank 1, traversing the tank 1 and the exchanger before
entering in the upper part of the tank 2. The flow is then distributed via two needle valves
in two different lines. The first valve allows to regulate the temperature between 4 and 250 K, the line being
wound around the joule heater's support located in the lower part of the cryostat's tube extension.
The second valve allows to reach temperatures as low as 1.5\un{K}, this line being extended
by a long and thin capillary (1\un{m} length and 0.5\un{mm} inner diameter) twisted inside the
cryostat's tube extension and ending with a final impedance (capillary of 5\un{cm} length and 0.15\un{mm} inner diameter).

The sample loading system is a tube of 7.6\un{mm} inner diameter and
0.2\un{mm} thick going through the whole length of the insert (from the
top at room temperature down to the vicinity of the magnetic field
center).  The sample zone is located at the center of the magnetic
field. It consists of a capsule of 800\un{\mu m} thick Torlon, a polymer
fairly transparent to x-ray, glued (Stycast 2850~FT) to the bottom end
of the sample loading tube.  A brass foil tube, radially drilled and
thermally anchored to the cryostat tube extension, surrounds the
capsule and is covered with several layers of super insulation
to screen against the thermal radiation from the Kapton windows at room
temperature.  The sample zone and the loading tube were designed to be
easily evacuated, then loaded with a few cm$^3$ of $^4$He exchange gas
at room temperature and pressure. In operation the external surface
of the loading tube is in direct contact with the liquid $^4$He of the
cryostat tube extension while its internal surface is in contact with
the exchange gas. The sample is thus cooled down by exchange gas to
the bath temperature.

To mechanically decouple the $^4$He insert from
the magnet and LN2 cryostat, and to allow fine adjustment of the $^4$He
cryostat's tube extension inside the vacuum enclosure, an additional
external support system (Fig.~\ref{Cryo3D}) was built
in order to avoid mechanical and thermal contact.

\begin{figure}[h]
{\centerline{\includegraphics[width=\columnwidth,clip,angle=0]{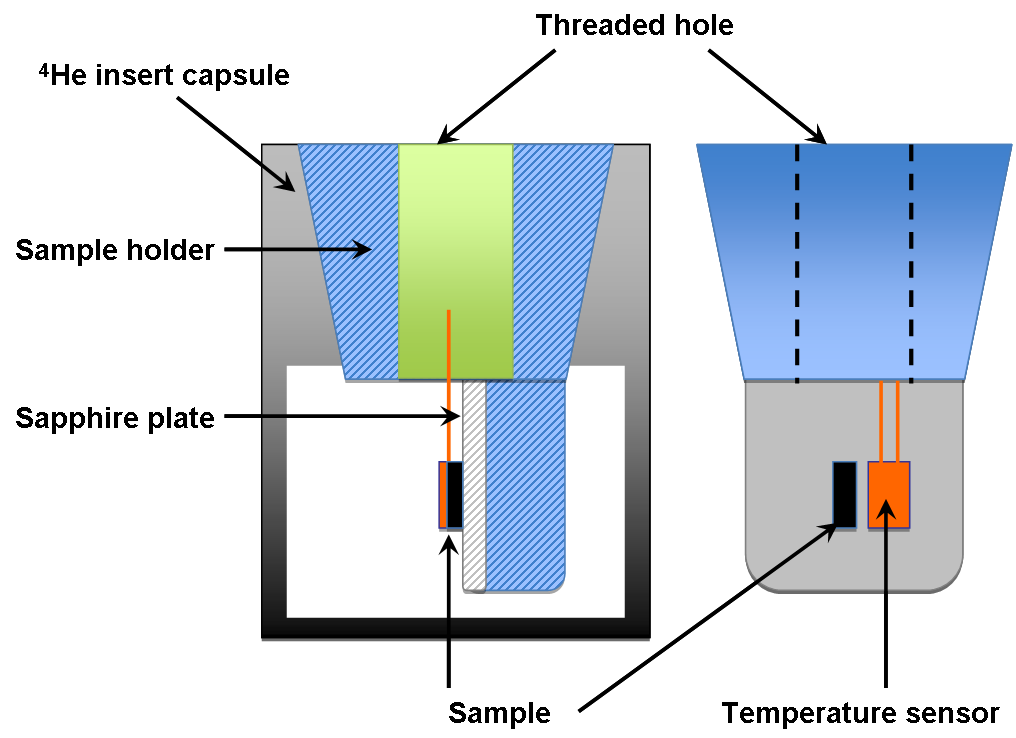}}}
\caption[Sample holder] {\small\sl{
    (Color online) Schematic views of the sample holder: (left) cross
    section of the sample holder inserted inside the capsule of the
    $^4$He insert, (right) front view. The temperature sensor is glued
    close to the sample and connected to a Lakeshore temperature
    controller via the measurement stick screwed into the threaded
    hole. $^4$He exchange gas flows around the sample thanks to holes
    drilled in the measurement stick and sample holder (not shown
    here).
}}\label{Sampleholder}
\end{figure}

Finally, a sample handling system has been designed specifically to
prevent sample motion (rotation and/or translation of sample holder)
induced by field pulses. It consists of the sample
holder coupled with the capsule located at the bottom of the $^4$He
insert. Both parts are made in Torlon, thus avoiding induced currents
in the sample holder. The upper parts of both elements
have a conical shape (see Fig.~\ref{Sampleholder})
allowing to physically lock the sample holder inside the $^4$He
cryostat after alignment of the sample in Bragg condition. The
stability of the sample orientation is thus ensured over time.  The
sample handling system is complemented with a measurement stick which
is equipped with
a pick-up coil and a heating resistance.  To improve thermal conductivity, a
thin sapphire plate (thickness~$\approx 400\un{\mu m}$) covers the part
of the sample holder where the sample is glued with GE varnish.

All the cryogenic environment has been custom designed and built at
the LNCMI-T.

\subsection{\label{sec:level2}Generator}

The split-pair magnet was used with a 24\un{kV}, 1.15\un{MJ}
transportable pulsed field generator developed by the LNCMI-T.  This
generator consists of two identical storage units
and one control unit. Each storage unit contains eight high energy
density capacitors of 250\un{\mu F} each, resulting in a total
capacitance of 4\un{mF}, two crowbar resistors ($R_{crowbar}(total)=
2\un{\Omega}$) and two inductive current limiters ($L(total)=
1.845\un{mH}$). The control unit houses the charger, thyristor stack,
dump resistors, the current and voltage monitors.  This unit is also
equipped with high-voltage, high-current pneumatically driven
commutators to give the possibility to work with two polarities.
Commutation and charging take only a few minutes and is controlled
completely by a computer integrated in the central control unit.  The
energy of the capacitor bank is released into the coil through an
optically triggered thyristor switch. The bank is protected from
failure in case of a short in one capacitor. To avoid an electrical dump
of all the other capacitors into the defective
one, they are all equipped with industrial HV-fuses.

The three subunits are mounted in individual steel frames so that they can be transported
using a fork lift or an overhead crane. Each storage module weighs
approximately 2000\un{kg} and has dimensions ($h \times d \times w$)
$2.00 \times 1.60 \times 1.20\un{m^{3}}$. The control unit is slightly
smaller ($1.50 \times 1.60 \times 1.20\un{m^{3}}$) and weighs only
800\un{kg}. The connections between the three subunits are made
on external patch panels and do not require opening of the units' enclosures.
The generator can thus be safely installed by users with minimal training.
For our synchrotron experiments, the generator, located
outside the experimental hutch, was linked to the personal safety
system (PSS) of the radiation hutch such that it could be charged only
after the hutch had been searched and interlocked.

\section{Commissioning}
\subsection{\label{sec:level2}Coil operation}

The coil parameters as measured during coil testing were: a resistance
value $R_{coil} = 192\un{m\Omega}$ and 1650\un{m\Omega} at 77 and
300\un{K}, respectively, an inductance $L = 35\un{mH}$ and a field
factor of 169.2\un{A/T}.

\begin{figure}[h]
{\centerline{\includegraphics[width=0.9\columnwidth,clip,angle=0]{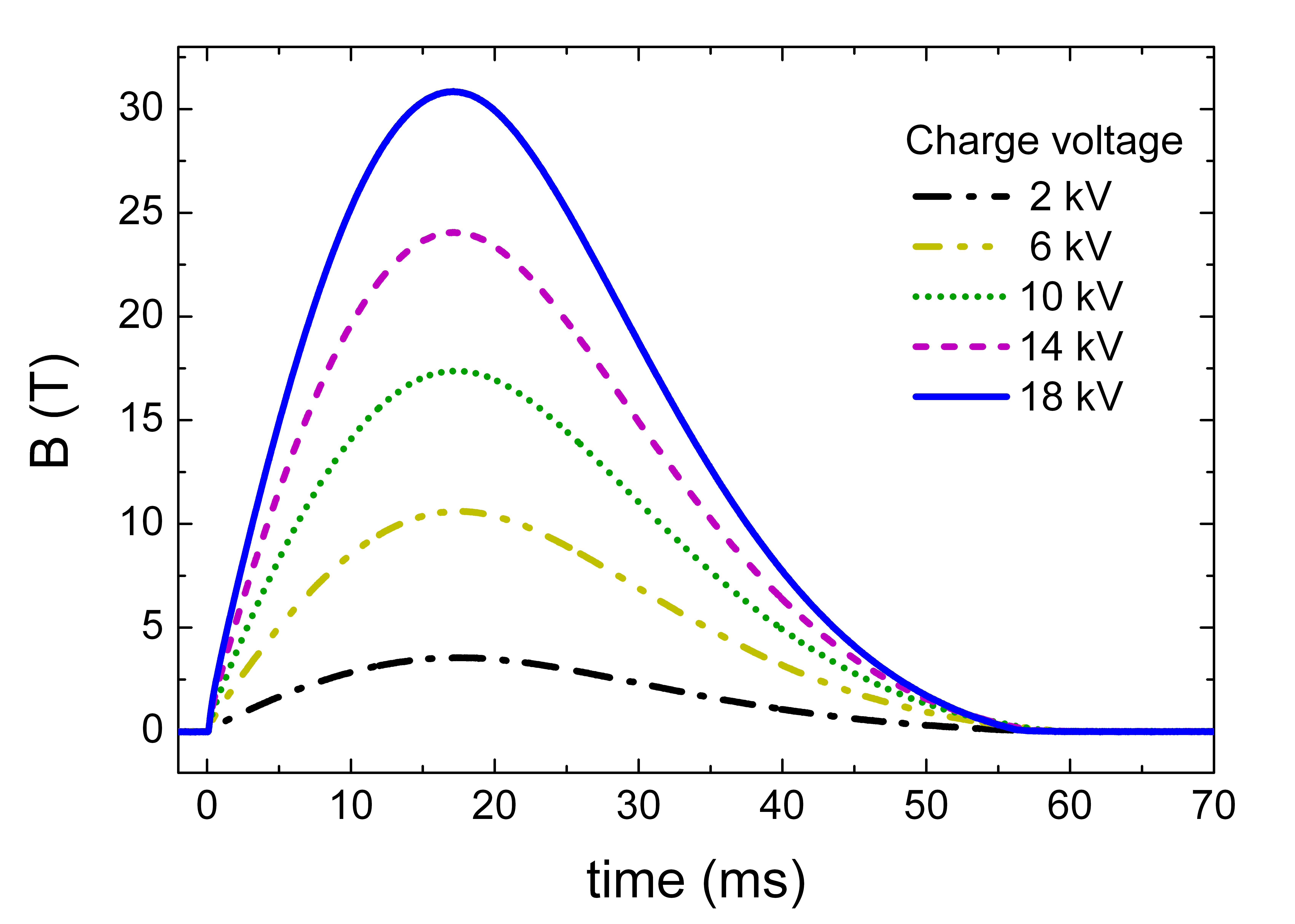}}}
\caption[Magnetic field pulse] {\small\sl{
    (Color online) Pulse field shape as measured at different charge voltages.
}}\label{Btracesnew}
\end{figure}
Maximum magnetic field of 31\un{T} was achieved at the center of the
split-pair magnet for a charging voltage of $V=18\un{kV}$.  The pulse
duration was of 60\un{ms} with a rise time of 16.5\un{ms}, as shown on
figure~\ref{Btracesnew}. With more than 4.5\un{ms} above 98\% of the
maximum field value, this pulse is relatively long compared to the
other pulsed field devices developed for x-ray and neutron scattering
experiments \cite{Matsuda2004, Narumi2006a, Narumi2006, Matsuda2006,
Linden2008, Yoshii2009, Islam2009, Islam2012}. In addition,
the magnetic field produced at the center of the magnet is fully
homogeneous over typical size of single crystals probed by x-ray
(i.e., from hundreds of micrometers to one or two millimeters) with
less than 1\% of variation over $\pm$3\un{mm} along the bore of the
magnet and less than 0.5\% radially (over a radius $R=3\un{mm}$). 
The energy stored into the coil at maximum field value is
relatively high ($\sim 480\un{kJ}$) and at the end of the pulse
its electrical resistance shows an increase of about $250\un{m\Omega}$
equivalent to a coil heating of 40\un{K}.
Around half an hour is then required between two pulses at 31\un{T}
to return to the coil operating temperature of 90\un{K}.

\subsection{\label{cryo}Low temperature measurements}
The cryogenic performances have been tested successively in open cycle
and closed cycle mode. In open cycle mode tank~2 is filled with
liquid $^4$He at 4.2\un{K} by opening the cold valve between tanks~1
and 2. The cold valve is then closed, and by pumping on tank~2
a sample temperature of 1.5\un{K} can be obtained within a few minutes.
This temperature can be maintained for around three hours before tank~2
has to be refilled.
\begin{figure}[h]
{\centerline{\includegraphics[width=\columnwidth,clip,angle=0]{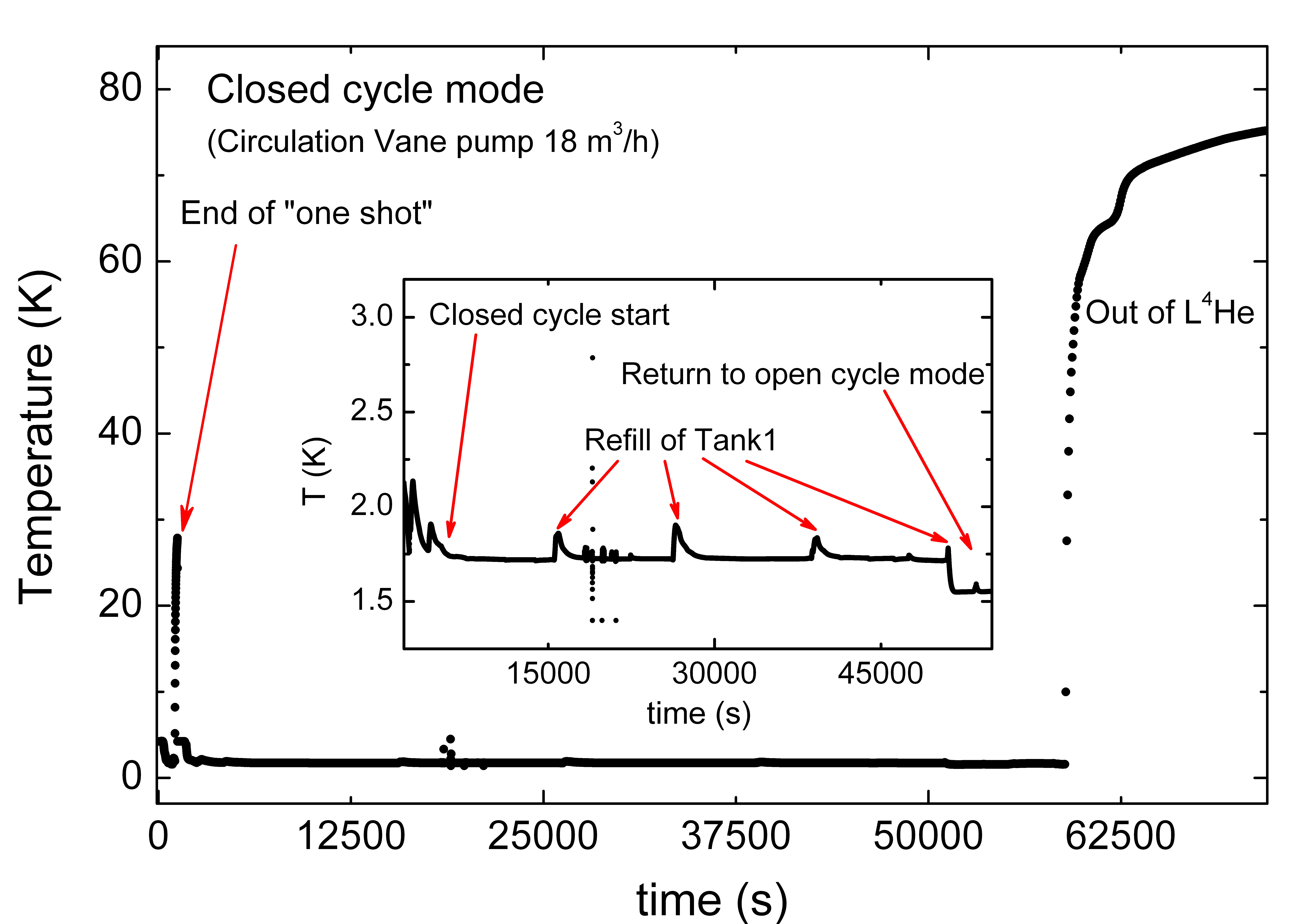}}}
\caption[Data acquisition scheme] {\small\sl{
(Color online) Monitoring of temperature when the $^4$He insert
    cryostat is used in closed cycle mode. The working temperature was
    measured with a CX-1050-SD Cernox sensor, calibrated down to
    1.4\un{K}. To obtain 1.5\un{K}, the Edwards E2M18 Hermetic
    circulation pump was replaced by a E2M40 pump.
}}\label{Closedcycle}
\end{figure}

In closed cycle mode, a continuous flow of cold He gas arrives close to the sample zone
through the capillary that traverses the tank~1,
the 2.5\un{K} heat exchanger, the tank~2, the cold needle valve and the final capillary,
whereas only the bottom part of tank 2 (tube extension) is filled with a few cm$^3$ of liquid $^4$He.
By pumping on tank~2 with a rotary vane pump (Edwards E2M40), a base temperature of 1.5\un{K} (1.7\un{K} when using a E2M18 pump, as shown on Fig.~\ref{Closedcycle}) is reached within less
than 20 minutes and can be maintained for more than 13 hours provided that tank~1 is refilled periodically.
In experiment conditions, i.e.~when the sample is exposed to x-ray,
this temperature can be preserved by using absorber foils before the
sample in the incident beam path, thus reducing the flux incident on the sample to $\sim2.10^{12}$ photons/s. For the Fe$_{1.1}$Te experiments shown here, at 31\un{keV}, we used two Sn foils of 50~$\mu$m each. In addition, a fast x-ray shutter (LS200 Laser shutter  from NM Laser products, Inc. (San Jose, CA, USA)) is used to expose the sample during very short time (100 ms per magnetic field pulse).

\section{High resolution X-ray single crystal diffraction and pulsed magnetic field}

Commissioning tests of the split-pair magnet were performed on the
undulator beamline ID06 (ESRF, Grenoble, France).

The source device was a cryogenically cooled permanent magnet
undulator (CPMU) \cite{Chavanne2007} with 18\un{mm} period and 6\un{mm}
minimum gap, such that the first harmonic could be tuned from
10.5\un{keV} to 19.0\un{keV}. A liquid nitrogen cooled Si (111) double
crystal monochromator selected photons with a relative band width of
$\approx 1.4\cdot 10^{-4}$. The maximum achievable photon flux was
about $2\cdot 10^{13} \un{photons/s}$ at 11\un{keV}.

The cryostat and magnet assembly was installed in the experimental
hutch 2 (EH2) on a high-load six circle diffractometer identical to
that of ID20 \cite{Paolasini2007}. The generator was installed in an
enclosure outside of the experimental hutch, and connected to the
magnet via a coaxial cable. A pneumatically operated switch
disconnected the generator from the magnet when the hutch was not
interlocked, i.e.~when access to the hutch was allowed.

\subsection{\label{sec:level2}Characterization of vibrations and angular stability}

\begin{figure}[h]
{\centerline{\includegraphics[width=\columnwidth,clip,angle=0]{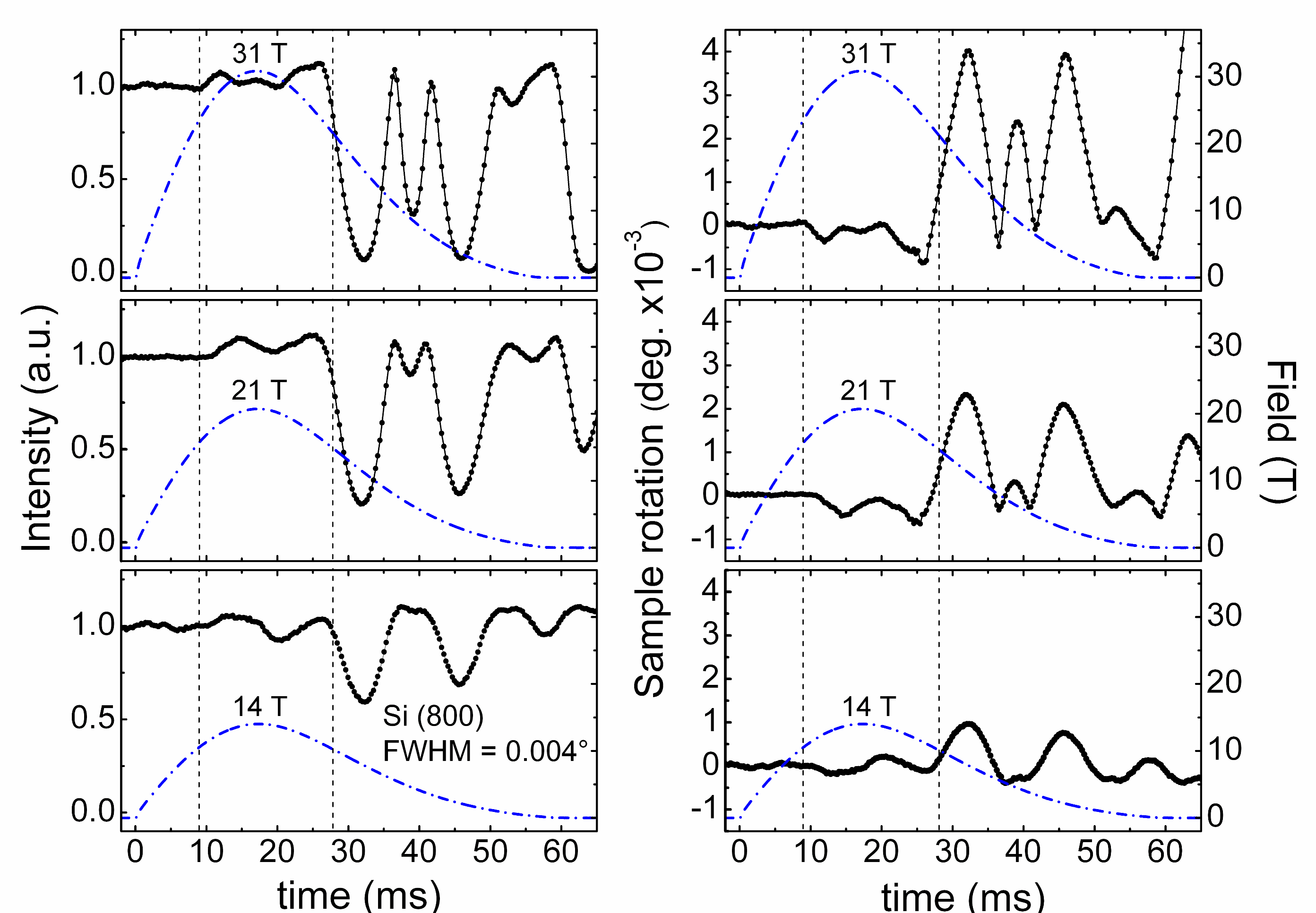}}}
\caption[Sample rotation versus magnetic field] {\small\sl{
    (Color online) (Left) Evolution of normalized intensity of a
    Si (800) Bragg reflection (FWHM($\theta$) $\approx 0.004^{\circ}$)
    as a function of time for different magnetic field pulses.  The
    pulsed field traces are represented by the dash-dot lines.
    (Right) Corresponding sample rotation as a function
    of time. Incident x-ray energy was $E=31\un{keV}$ ($\lambda=$
    0.3995\,{\AA}).
}}\label{IntandRotnewc}
\end{figure}

The tests were first carried out on silicon single crystals to
estimate the vibration amplitude of the complete device.  The
diffracted intensities were measured with a silicon photodiode mounted
on the detector arm of the diffractometer and recorded (along with
generator voltage and current, and the voltage on a pick-up coil on
the sample stick) during the field pulses on a digital data recorder
(HIOKI Memory HiCorder model 8860). Motorized slits upstream of the photodiode
detector at 70\un{cm} from the sample position were used
to adjust the longitudinal ($2\theta$) resolution. Due to the very
narrow rocking curve width of the silicon crystal and the low
divergence of the incident x-ray beam this configuration is extremely
sensitive to rotations of the sample during the field pulse.
	
One main source of vibrations identified on the beamline was eddy
currents propagating in the aluminium parts of the diffractometer.  To
increase the distance from the field center to those parts and suppress
the repulsive forces generated by induced currents, several aluminium
parts such as the goniometer top plate and rotation stages ($\chi$ and
$\varphi$ stages) of the diffractometer were removed and replaced by
fiber glass spacers (FR4/G10).  The remaining aluminium parts were
thus located at more than 49\un{cm} from the center of the magnet.
Induced eddy currents and corresponding forces were
sufficiently weak to not disturb the measurement anymore.

As an example, the evolution of the normalized intensity of a Si (800)
Bragg reflection (FWHM($\theta$)$\approx0.004^{\circ}$) is shown on
figure~\ref{IntandRotnewc} (left) as a function of time for different
amplitudes of field pulses: 14, 21 and 31\un{T}.  The intensity begins
to oscillate 9\un{ms} after the magnetic field pulse was
triggered. However, the variation of intensity does not exceed 17\%
during the first half of the pulse (i.e.~up to 28\un{ms} after the
beginning of the pulse).  Then, between 28 and 60\un{ms}, the
amplitude of the oscillation increases, with the intensity even going
through zero for the 31\un{T} pulse.  The orientation of the sample is
nevertheless maintained since the intensity is recovered with no more oscillation
1\un{s} after the pulse (not shown).

By measuring the change of intensity at slightly lower or higher
sample angles ($\theta$) than the Bragg position, it was
possible to ascribe the loss of intensity to sample rotation. The
latter was easily calculated as a function of time based on the full
width at half maximum of the reflection and pseudo-Voigt like shape of
Bragg peaks. The result is displayed on figure~\ref{IntandRotnewc}
(right).  These curves show that sample rotation is always lower than
$\pm0.001^{\circ}$ during the first half of the pulse and does not
exceed $\pm0.0005^{\circ}$ at maximum field, well below the mosaicity
of most single crystal samples of interest for high-field studies
(e.g.~$>0.05^{\circ}$ for typical intermetallic compounds).  In the
second half of the pulse, the rotation is as high as $0.004^{\circ}$
for the shot at maximum field value.

This means that our experimental setup is not suitable for studying
intensity changes over the entire field pulse for very high quality
samples with mosaicity below $0.05^\circ$. One should note, however,
that the usable pulse length
is at least five times higher than the one provided by the other
pulsed field devices dedicated to x-ray single crystal diffraction
experiments \cite{Narumi2006a, Narumi2006, Islam2009, Islam2012}.
This is of advantage for studying metallic samples and hysteresis
effects. Furthermore, the longer rise time induces less sample
heating due to induced currents.

Other efforts have been carried out to further suppress or reduce vibrations,
but they yielded no significant improvement, e.g. decoupling the
coil cryostat from the diffractometer and sample cryostat by means of
viscoelastic absorbers, and increasing the stiffness of the
diffractometer by means of external reinforcements.

While simple, cheap and easy to use, a point detector such as
a photodiode requires at least one magnetic field pulse per scattering
angle, and is thus not exactly very efficient for studies of magnetic
field induced crystallographic phase transitions.
On the other hand, a fast 2D detector can follow the evolution of Bragg
peaks during magnetic field pulses \cite{Narumi2006a, Narumi2006},
i.e. it allows the observations of intensity changes at several
scattering angles simultaneously. In our
first experiments a MAXIPIX detector \cite{Ponchut2011} was used. This
is a fast readout, active pixel detector based on the Medipix2
\cite{Llopart2002} and Timepix \cite{Llopart2007} readout chips
developed by CERN and the Medipix2 collaboration. The system achieves
up to 1.4\un{kHz} frame rate with 290\un{\mu s} minimum readout dead
time and has a pixel size of $55 \times 55\un{\mu m^2}$. We used the
single chip ($256 \times 256$ pixels) detection geometry. In this
geometry, with the detector mounted on the detector arm ($\gamma$-arm) of the
diffractometer at a distance of 53\un{cm} from the sample, the accessible angular width
in both the vertical and horizontal direction is around $\pm$\un{0.8}$^{\circ}$,
and the angular resolution corresponding to the pixel size is approximately
0.006$^{\circ}$ in 2$\theta$. The resolution could be increased by
increasing the distance to the sample, e.g. to 0.0037$^{\circ}$ at
85\un{cm} from the sample. This, however, reduces the angular field of view to
$\pm$\un{0.5}$^{\circ}$ in both the vertical and horizontal direction.

\begin{figure}[h]
{\centerline{\includegraphics[width=\columnwidth,clip,angle=0]{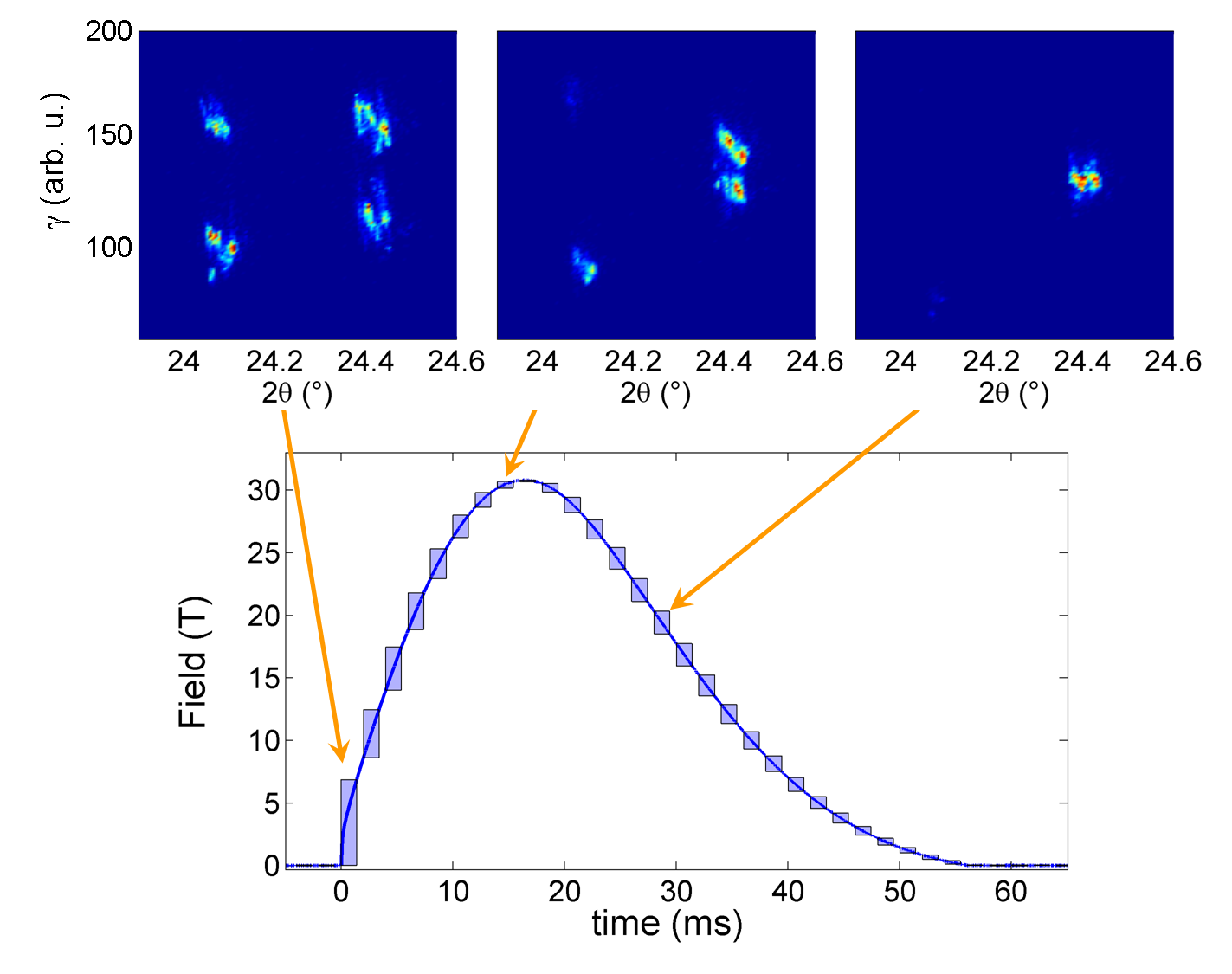}}}
\caption[Data acquisition scheme] {\small\sl{
(Color online) Data acquisition scheme used combining the split-pair magnet system and a MAXIPIX detector.
Line: trace of magnetic field pulse; boxes: acquisition windows. Each frame was recorded every 2~ms,
with an exposure time of 1.4~ms and a readout dead time of 0.6~ms.
Incident x-ray energy was $E=31\un{keV}$ ($\lambda=$~0.3995\,{\AA}).
Distance sample-detector $\approx$ 53\un{cm}.
Some selected frames above the pulse trace present the field dependence of
the (400)/(040) reflections of a Fe$_{1.1}$Te crystal.
The temperature of the sample was T~=~23\un{K}.
}}\label{Dataframe}
\end{figure}

The high frame rate of this detector perfectly matches the time structure
of magnetic field pulse and enables multi-frame acquisition
as already implemented in other x-ray techniques combined with pulsed fields
\cite{Strohm2011}. Similarly, synchronized
with the field pulse, the MAXIPIX detector allows to follow the evolution
of Bragg reflections as a function of time during a magnetic field pulse.
In our experiments, each frame was exposed during 1.4~ms with a data
transfer time of 0.6~ms resulted in an effective frame rate of 500\un{Hz}.
In this multi-frame acquisition mode, 50 frames were acquired for
each magnetic field pulse, with two frames being recorded before
the triggering of the pulse (see Fig.~\ref{Dataframe}).

Raw data obtained are images of the reciprocal space with the scattering angle 2$\theta$
along the horizontal direction. For the following, we will call
$\gamma$ the out of diffraction plane angle (vertical direction on raw images)
by analogy to the detector arm.
To extract the diffracted intensity as a function of time, frames measured
in a single magnetic field pulse are first $\gamma$-integrated. The resulting 2$\theta$-scans are then
concatenated to form an image showing the diffracted intensity at specified scattering angles
as a function of time. Typical plots obtained are presented figure \ref{FeTe_H_T}.
From these plots, it is then easy to get the intensity variations as a
function of field.

\subsection{\label{sec:level2}Field-induced phase transition in Fe$_{1.1}$Te}

First experiments were performed on Fe$_{1.1}$Te single crystals. This compound belongs to
the parent phases Fe$_{1+x}$Te of chalcogenide superconductors Fe(Te,Se),
which are the subject of intense studies because of
their similarities and differences with high T$_c$ cuprates. At high temperatures,
Fe$_{1+x}$Te compounds crystallize in the tetragonal \textit{P}4/\textit{nmm} structure
\cite{Bao2009}. Upon cooling the temperature through $T_N \simeq$~57-70\un{K}
\cite{Johnston2010, Bao2009, Li2009, Rodriguez2011, Rossler2011},
they undergo a first order phase transition leading to an antiferromagnetic (AFM)
ordering associated with a structural distortion
lowering the high temperature tetragonal paramagnetic lattice
symmetry. For $x<$~0.12, the low temperature phase is monoclinic
\textit{P}2$_1$/\textit{m} with a bicollinear AFM order whereas
it is orthorhombic \textit{P}\textit{nmm} with a strongly incommensurate helimagnetic
spin order for $x>$~0.12 \cite{Bao2009, Li2009, Martinelli2010, Rodriguez2011}.
In these systems, magneto-elastic couplings are suspected to play a crucial role
on their magnetic and electronic properties \cite{Paul2011,Tokunaga2012}.
Recently, an irreversible alignment of the AFM moments inducing a step-like anomaly
in the magnetization was reported at low temperature and high field for
Fe$_{1.1}$Te single crystals \cite{Knafo2013}.
The critical field associated with this transition was denoted $H_R$.
A spin-flop-like reorientation of the AFM  moments was suggested to explain this behavior,
keeping in mind that the microscopic model should also describe
the remanent moment reorientation once a magnetic field higher than $H_R$ is applied.
To probe magneto-elastic effects that could be the cause of this behavior,
we performed diffraction experiments on Fe$_{1.1}$Te single crystals over
the temperature range $1.6<T<$~70\un{K} and in pulsed magnetic fields up to 31\un{T}.
The detailed study of this system will be published elsewhere \cite{Fabreges2014}.
Here, only first results obtained illustrating the possibilities offered by
our pulsed field system as regards of field, temperature and stability will
be shown.

Single crystals of Fe$_{1.1}$Te, grown by a modified Bridgman method, were kindly provided
by Viennois and Giannini from Geneva University (Switzerland). A single crystal sample
was mounted on the sample holder as described above.
The sample was aligned with [1~0~0] and [0~0~1] axes in the diffraction plane giving access
to Bragg reflections in the (H,~0,~L) zone axes. The magnetic field was applied along
the \textit{\textbf{b}}-axis of the crystal.
X-ray diffraction images were recorded as described above, using an x-ray photon energy of 31~keV.

We first performed zero field measurements to collect the temperature
dependence of several Bragg reflections at different scattering angles.
The high temperature tetragonal (400) and (404) Bragg peaks were, for instance, measured.
These measurements confirm that our Fe$_{1.1}$Te sample
undergoes a structural transition from a tetragonal
lattice symmetry at room temperature to a monoclinic one below
$T_N \simeq$~58\un{K}, in agreement with previous results \cite{Rossler2011, Knafo2013}.
Our observations are consistent with the previously determined
monoclinic lattice parameters, $a=$~3.83378(6)~{\AA}, $b=$~3.78667(8)~{\AA}, $c=$~6.246427(8)~{\AA}
and  $\beta=$~89.359(1)$^\circ$.
The inserts in figure \ref{Dataframe} show typical data acquired during a
single magnetic field pulse at $T=$~23\un{K}, below the phase transition.
The leftmost image, acquired at $B=$~0 before the field pulse, clearly shows
four reflections corresponding to the twinned monoclinic domains that
arise at the structural phase transition \cite{Tokunaga2012}.
The peaks are identified as (400) at 2$\theta=$24.10$^\circ$ and
(040) at 2$\theta=$24.40$^\circ$.

Next, we applied the magnetic field.
The evolution at $T=$~23\un{K} of the (400)/(040) Bragg reflections during a 31\un{T} pulse
is illustrated, for instance, by the three selected frames shown on figure \ref{Dataframe}-top.
At the beginning of the pulse, four Bragg peaks are observed, as described above.
Then, during the rise of the field, a progressive disappearance of the low angle Bragg peaks
at the benefit of the high angle ones occurs. Finally only one peak remains
at the end of the pulse as shown on the frame
measured during the fall of the pulse (Fig.~\ref{Dataframe}, right frame).
This clearly reveals a magneto-crystalline selection of domain leading to the detwinning of the crystal.
This behavior was observed for all temperatures in the range $20<T<$~50\un{K}, the critical
field required to completely detwin the crystal increasing when temperature decreases.

\begin{figure}[h]
{\centerline{\includegraphics[width=0.8\columnwidth,clip,angle=0]{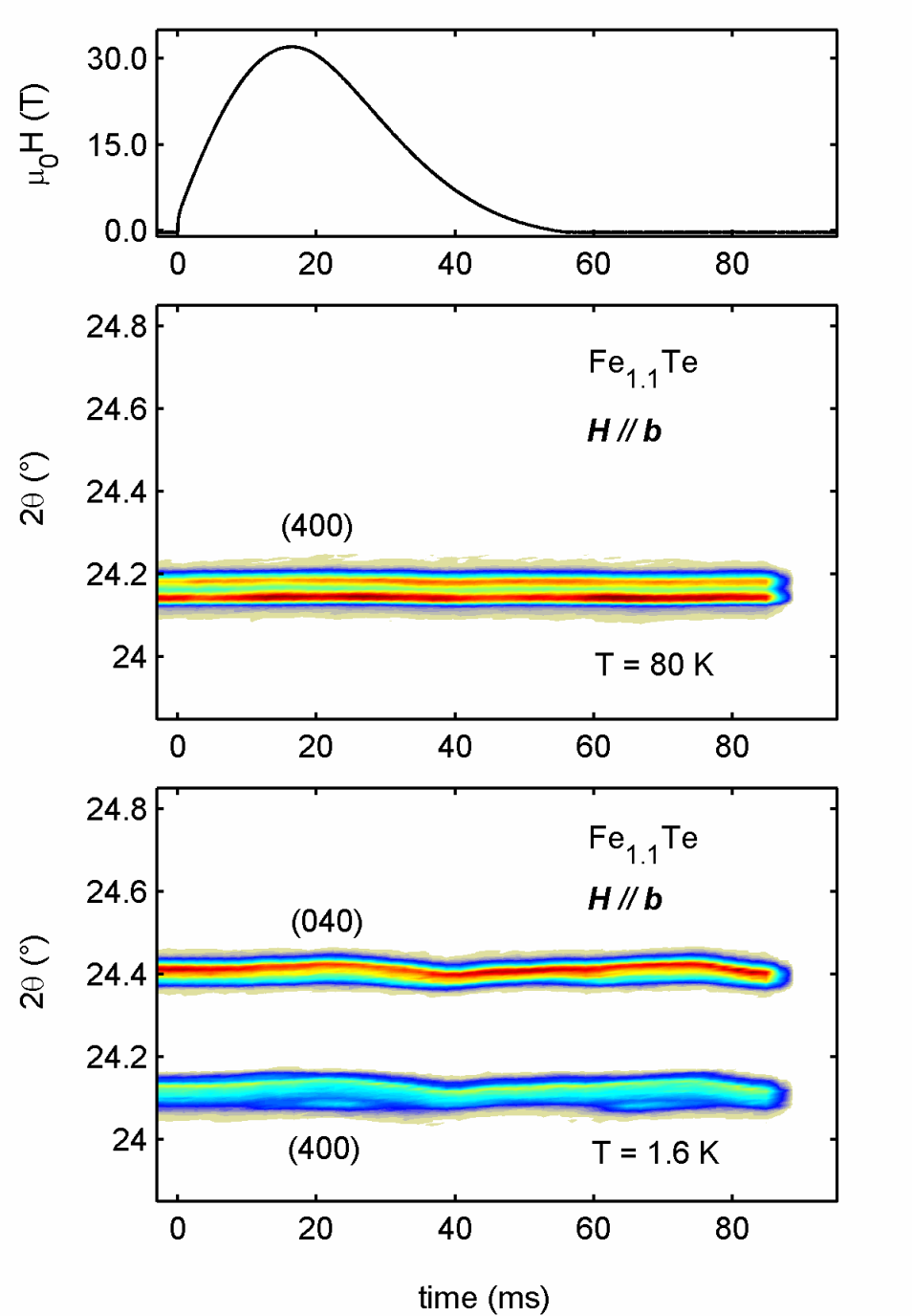}}}
\caption[Measure] {\small\sl{
  (Color online) Time evolution of $\gamma$-integrated intensity
  of the (400) Bragg reflections of Fe$_{1.1}$Te at different
  temperatures when a magnetic field pulse of 31\un{T} is applied.
  Below T$_N=$\un58\un{K}, the tetragonal (400) splits
  in the (400)/(040) pair of reflections because of the monoclinic
  distortion. During the magnetic field pulse (between 0 and
  60\un{ms}), there is nearly no evolution of the integrated intensity
  indicating a quite good sample stability with regard to the beam.
}}\label{FeTe_H_T}
\end{figure}

Figure~\ref{FeTe_H_T}-bottom shows the time dependence of the integrated intensities of the
(400)/(040) pair of reflections measured at $T=$~1.6\un{K} when a 31\un{T} magnetic field pulse
was applied. Only slight oscillations are visible, both peaks keeping their intensity
during all the pulse. This clearly indicates that the high field limit
of our experimental set-up is not enough to detwin this compound at low temperature.
This behavior is in agreement with high field magnetization measurements \cite{Knafo2013}
and will be further explained elsewhere \cite{Fabreges2014}.

Above the N\'{e}el temperature (Fig.~\ref{FeTe_H_T}, middle)
no evolution of the integrated intensity of the different Bragg peaks measured
was detected as a function of time (and therefore as a function of field),
clearly indicating a quite good sample stability with regard to the beam during a pulse.

This first experiment thus shows that applying high magnetic field results in
magneto-crystalline domain selection in Fe$_{1.1}$Te. This behavior
can be directly followed during the entire magnetic
field pulse thanks to the
multi-frame acquisition mode allowed by the fast two-dimensional pixel detector
associated with the pulsed field set-up developed for single crystal x-ray diffraction.
This promising results open the door to the investigation of the $H$-$T$ phase
diagram of this material in the limits offered by our experimental set-up.

\section{Conclusion}
In summary, we have constructed a split-pair magnet implemented in a novel cryogenic environment
to perform high field single crystal x-ray diffraction measurements
at synchrotron sources. This new device offers a panoramic access to x-rays
allowing measurements at any scattering angle. It provides
magnetic fields up to 31\un{T} with a long pulse duration (rise time of 16.5\un{ms}),
which is of great advantage for studying metallic samples and hysteretic effects.
Furthermore, a temperature as low as 1.5\un{K} can be maintained on the sample
for more than ten hours using the closed cycle mode of the $^4$He cryostat
inserted into the bore of the magnet.

During the commissioning of this apparatus, particular attention was given
to the problem of field-pulse induced vibrations of the sample.
We succeeded to limit the sample rotation induced by magnetic field pulse
to $\pm 0.001^{\circ}$ during the first half of the pulse. This rotation
is well below the mosaicity of most single crystal samples
of interest for high field studies.

Combined with a fast two-dimensional pixel detector adapted to the time structure
of the field pulse, it offers the possibility to follow the intensity changes
of several Bragg peaks at different scattering angles simultaneously and throughout
the whole duration of the pulse.

Improvements could be carried out on the duty cycle of the experiment
by incorporating LN2 cooling channels in the design of the coil.
This will decrease the cooling time of the coil but this will come
at the expense of the maximal field or will require higher energy.

This device offers the opportunity to investigate small crystallographic changes,
such as magneto-elastic effects, field induced symmetry breaking or charge order
that could occur in strongly correlated electrons systems
like, e.g., high $T_c$ superconductors, quantum spin systems or heavy fermions
materials.

\begin{acknowledgments}
The authors would like to thank F. Lecouturier and N. Ferreira of the LNCMI-T/CNRS for the characterization
of the wire conductor (resistivity and wire strength measurements). J.-M. Lagarrigue and L. Bendichou who machined
the steel spacers and other pieces required for the coil building are also greatly acknowledged.

At the ESRF, we would like to thank H.-P. Van der Kleij for lending us his autocollimator,
P. Van Vaerenbergh and P. Bernard for the different mounting systems developed to reduce vibrations,
and P. van der Linden for his help with cryogenics.

The authors acknowledge the ESRF for granting the beam time for these developments and experiments.
The safety issues of the equipment were addressed with the help of ESRF Safety Group.
The MAXIPIX photon-counting pixel detector was lent to the authors by the ESRF detector pool.
Part of this research was funded by the ANR (Grants N$\circ$. ANR-05-BLAN-0238 and ANR-10-BLAN-0431).
\end{acknowledgments}

%

\end{document}